# Ten Quick Tips for Using a Raspberry Pi

Anthony C Fletcher[1]* and Cameron Mura[2]

[1]Salix Management Consultants Ltd, London, United Kingdom;
[2]Dept of Biomedical Engineering; University of Virginia; Charlottesville, VA 22908, USA
journal: *PLOS Computational Biology* "Quick Tips" format; draft of 06-Feb-2019
*correspondence: drtonyfletcher@gmail.com

## Overview

Much of biology (and, indeed, all of science) is becoming increasingly computational. We tend to think of this in regards to algorithmic approaches and software tools, as well as increased computing power. There has also been a shift towards slicker, packaged solutions—which mirrors everyday life, from smart phones to smart homes. As a result, it's all too easy to be detached from the fundamental elements that power these changes, and to see solutions as "black boxes". The major goal of this piece is to use the example of the Raspberry Pi—a small, general-purpose computer—as the central component in a highly developed ecosystem that brings together elements like external hardware, sensors and controllers, state-of-the-art programming practices, and basic electronics and physics, all in an approachable and useful way. External devices and inputs are easily connected to the Pi, and it can, in turn, control attached devices very simply. So whether you want to use it to manage laboratory equipment, sample the environment, teach bioinformatics, control your home security or make a model lunar lander, it's all built from the same basic principles. To quote Richard Feynman, "*What I cannot create, I do not understand*".

**Introduction**

Recent Raspberry Pi sales figures [1] show that this humble board is the world's third best-selling general-purpose computer, with over 15m sold (after PCs and Macs). While it may be a fair bet that many of them are sitting unused in dusty cupboards, equally many of them are beavering away at a multitude of practical tasks, at home and work, day-in, day-out, without blinking (They do blink, they have an LED!). One of us (ACF) has one he hasn't rebooted or touched in over a year. It just does its work. What is the Pi, and why should anyone care?

For those of you new to this realm, the Pi is a small computer made and sold by the trading arm of a UK-based charity [2] whose initial aim was to "promote the study of computer science and related topics, especially at school level, and to put the fun back into learning computing". A historical account of the Raspberry Pi Foundation's early roots is offered in [3]. The Pi's integrated circuit design is that of a System–on–Chip (SoC), versus the usual motherboard-based PCs familiar to most readers. A major architectural difference here is that the core components in an SoC-based microcomputer—e.g., the CPU-housing microcontroller, memory blocks, voltage regulators, etc.—are built as a single, fully integrated circuit, enabling small chips with low power consumption, reduced heat dissipation, and so on. These features, in turn, have made such systems popular for embedded and mobile devices. The Pi uses an ARM chip, such as those found in about half of all smartphones [4] (as opposed to the Intel/AMD chips found in most PCs). About as powerful as an iPhone 5 [5], the Pi generally runs a special distribution of the Debian (Unix-based) operating system (OS), termed 'Raspbian' (Figure 1); other Linux distributions, as well as other (non-Linux) OSes can be run, too.

In practice, the Pi is a small computer (without a case)—about the size of a bar of soap (Figure 1A). It features the usual USB ports and HDMI sockets, as well as one extra set of gizmos (which are what make the Pi so versatile), known as a set of "general purpose input-output" (GPIO) pins. The GPIO, comprised of 40 pins, enables you to directly tell the Pi to send or read a voltage on them. This, in turn, enables you to plug-in about a gazillion peripherals and read and control them directly, all without any soldering. Using a little device known as a 'breadboard' [6], you can also rapidly build and control basic electronic circuits (often used in prototyping electronic circuits, the breadboard is a device to link resistors, transistors and other components, without soldering).

This simplicity and modularity of the Pi is one of the reasons why it has become so popular: after a small learning curve, you aren't limited by your skills, only by your imagination. The Pi is a real jack of all trades and it is manifestly not (just) a toy: for example, Los Alamos National Laboratory recently built a Raspberry Pi cluster with 3,000 cores as a pilot before scaling up to 40,000 cores or more [7,8].

A Pi shouldn't be confused with, for example, an Arduino. The latter, which is an open-source, single-board microcontroller, does not run an OS; instead, a microcontroller is designed to run relatively simple programs over and over, depending on the intended use (e.g., as a temperature or light sensor), often as an embedded system. The Arduino is quite useful, but differs somewhat from the Pi in its range of applicability and intended uses; helpful comparisons of the Pi and Arduino can be found at [9,10].

So, the Pi is an inexpensive device—a mini-computer, basically—that can have a place in any lab, home, car, garden and office. Here are ten quick tips for using one. Note that these tips are intentionally broad in scope, so as to illustrate the general utility of this sort of device—including in the professional development of any scientist, by virtue of developing one's computational skill-set and knowledge of computer hardware principles.  As a more concrete context, in alignment with this journal's more typical readership, some explicitly biosciences-related examples are included in Tip 10.

**Tip 1: Don't be put off by the Pi being incredibly inexpensive** (in fact, there's an even cheaper version...)

For maybe $60 USD—or less, if you have items like an SD card or a suitable mobile phone charger lying around [11]—you have all that you need to get going.  The Pi could even pay for itself: just use one to monitor the energy use in your house and turn the thermostat down a bit occasionally.  And, if the kids knock a soda onto it, who cares—there are no moving parts.  A smaller-sized 'Pi Zero W' is also available, which comes with built-in WiFi and Bluetooth, and retails for about $10 USD (however, you will need, at a minimum, power and an SD card).  Many individuals use a full-size Pi, such as a third-generation Raspberry Pi 3, to prototype their application, and then simply run the production version on the smaller device (if the usual task/load isn't too compute-intensive).  The Pi Zero W is ideal for remote cameras and environment sensors, and can work for considerable periods supplied only by a mobile phone power pack.  After a "coffee incident", one of us (ACF) rinsed a Pi Zero W under a tap and it worked fine once it had dried. (It was unplugged, and this is not recommended, but c'est la vie.)

**Tip 2: Try it, if you're interested in more than just *using* a computer**

Many of us who are scientists probably remember the thrill of discovering things for the first time as children. The Pi gives you the chance to relive such moments as an adult. We mentioned earlier the small learning curve to get a Pi up and running. Small it is, but it's there and it's a challenge to overcome.  Once you decide what you need, a few Google searches and you have access to the collective knowledge of many such enthusiasts and 'maker' communities [12,13].  Therein lies its charm: while the Pi is (intentionally) not slick, and it doesn't "just work" (like turning on a tablet and firing up an app), its configurability is virtually unlimited.  You learn to do so much more with it—not just as an open-source PC, but as a "brain" to control a multitude of peripherals.

You can re-discover your old-school physics about what resistors, transistors and capacitors do and, even if you don't like coding, it's still relatively straightforward to use the Pi as there is a Graphical User Interface (GUI) and many pre-built applications. If you are so minded, learning the basics of two languages (PHP, Python) and a smattering of Unix/Linux can get the device really singing.  The Pi is like a kind of modern, digital, utilitarian jigsaw kit.  All the parts and code are out there, you can just put them together to make something bespoke; unlike a jigsaw, Pi-based devices are much more than just decorative.

**Tip 3: If you like repetitive work, don't buy one**

As with any computer, the Pi excels at automating tasks, and especially so, as it's simple to connect to a wide range sensors and external inputs. There's always a toss-up between writing a little script to automate something, versus just doing it manually (on occasion).  How you decide to proceed—to script or not to script—is generally dictated by your personal balance of two countervailing forces: (i) the effort required of you (the cost of scripting) versus (ii) the frequency with which the task arises (often, or only intermittently?—this is the cost of not scripting). If you tend to like the idea of scripting/programming and automating, then you'll probably enjoy the Pi from the get-go. Particularly with a shell script or perhaps Python, R or MATLAB, it becomes as quick as possible to use the Pi to automate processes. It's very easy to add the Pi to your home or lab network, share drives, etc. (all behind a firewall of course) and simple, one line, scripts that run at regular intervals ("cron jobs", after the standard Unix command 'cron') can be written for repetitive tasks, for example turning on the coffee pot at 6:00 each morning.

So, you can simply automate things that would be just too laborious (impossibly so) with a pen and paper, or are otherwise unamenable to manual intervention. For example, during a recent hot spell one of us (ACF) became interested in how the temperature in the house varied with the windows open and the blinds drawn. The measuring peripheral cost $5 and the code took about ten minutes. (Search your favourite auction site for a DHT22 or AM2302 temperature and humidity sensor, or a BME280, which adds barometric pressure as well.)

**Tip 4: Have fun**

Have you ever wanted to turn the lights on when it gets dark, without getting up? Have an RFID controller on a fridge to control who accesses it? See who is at your front door? Learn what is eating the nuts you left out for the squirrels? Shine a laser beam across the doorway so you can log who comes in and out? (And maybe take a photo or HD video? For around another $25 you can get a very decent camera.) A Pi opens the door to all of these capabilities—all in addition to tasks like logging, temperature recording, basic artificial intelligence (AI) and 'on-the-go' data science, image recognition, text-to-speech conversion and vice versa.

Essentially, you are limited only by what you want to do: from basic robotic kits to mechanical arms, there are a host of possibilities. And, many of these activities and projects have been done already (or at least started), so you can simply Google them and use other work or build on people's existing efforts (see also Tip 7).

Because of its simple interface with sensors, motors, cameras, etc., the Pi can be used for many practical tasks. Yesterday's youngsters loved model railways and constructor sets, but today they might love to make motion sensor-based teddy bears that laugh when you wave your hand.

Just one example: using MIT's App Inventor [14], it's easy to make a little app that, when you say "Play X" into an Android phone, it tells the Pi plugged into the hi-fi to do just that. Now no one in the house, reading the paper in silence, is safe from a high volume rendition of Bobby Boris Pickett and the Crypt Kicker Five blasting out Monster Mash.

**Tip 5: Embrace open-source**

Much of the software commonly used on a Pi is free of cost, and there are plenty of great tools to bring it all together. For example, the Pi's open-source Rasbian OS, mentioned above, is based on Debian, with all that that involves; indeed, much of the software and peripherals one uses with a Pi comply with free and open-source (FOSS) principles, meaning no restrictive licenses or costly fees. (For more on the issue of 'free', as in licensing & use ['freedom'] versus cost ['free pizza']—as well as the general issue software licensing and FOSS—see ref [15].)   It would be remiss of us not to expand on the exceptional MIT App Inventor tool.  This is a visual programming environment that allows anyone—even children—to build fully functional apps for smartphones and tablets.  It's brilliant with a Pi, as you can easily write basic "Apps" to interface your phone or tablet with your Pi and communicate over the internet or just your own local area network (or 'LAN'; basically, a home network). You can also, of course, run a basic web server on the Pi and communicate that way. It's as easy as Pi. The world really is your oyster here.

Although not Pi–specific, many organisations and data suppliers (especially public utilities) also offer application programming interfaces (APIs) that enable you to 'talk' to them and get their data in a structured way.  For example, one of ACF's Pi's automatically gets the local bus times and flashes up when the next bus is due, so he knows when to stroll down to the stop and not have to hang around.

And, in addition, you can give back to the community by sharing your code with others (which you're encouraged to do, fear or anxieties aside [16]!).  In Tip 7, we mention the repositories of data for the Pi; well, you can easily answer others questions here, or even set up your own simple web site to share your code (which of course can be hosted on your Pi!).  People will certainly appreciate any and all sharing.

> **Note**
>
> If the Journal thinks it is of interest, ACF is prepared to set up a basic website (see: tensimplepi.malvernskylarks.com) to answer direct questions from readers about how to get started and answer basic questions. ACF would undertake to oversee it for at least a year. It would be totally non-commercial.

**Tip 6: Welcome an IoT that isn't going to disappear**

The Internet of Things [8,17] is already upon us, and will only increase in its ubiquity; and, it's currently a standards nightmare.  From a smart fridge to smart heating (well, smart anything), off-the-shelf solutions are generally quite expensive and, more crucially, closed-source and proprietary.  So, your light bulbs might not talk to your central heating, and your front doorbell could struggle to communicate with your fridge.  More importantly, if the manufacturer goes broke you'll be left with a lot of useless consumer gear.  It doesn't have to be like this, though: Using open-source rather than proprietary technologies—e.g., open-source software libraries, modern programming languages and standards-compliant tools like UDP (User Datagram Protocol) and Rf (Radio frequency) communication for networking—you can easily communicate between devices and across LANs and the internet.  It's also simple to converse with virtual assistant devices such as Siri and Alexa via their APIs.  More generally, your Pi now places you into the realm of IoT concepts like 'edge' and 'fog' computing [17].

You can be as sophisticated or as simple as you like—with a Pi Zero W at $10 it doesn't break the bank to try things.  Everyone will have a different set of needs.  As well as some of the uses mentioned earlier, consider $3 wireless switches that tell if a door or window opens, or similarly priced passive infrared (PIR) sensors and a microwave sensor that you can wave at to turn on the mains power (relatively inexpensive at $6 a pop). One of ACF's next tasks is to

set up a dedicated barcode scanner by the kitchen waste bin to remind him what to buy at the supermarket based on what he throws away.

**Tip 7: Engage with online communities (they can be helpful and virtuous places, just beyond the trolls, fake news and Twitter bots)**

One of the great things about the Pi—and, in general, the sociology and culture that underlies Linux, Python, PHP and other open-source efforts [18]—is that the folks who contribute to these communities typically have a genuine interest and desire to both share information and help others [19]. There is generally much prompt and free help, and opportunities for collaboration may even arise. Indeed, large communities online are dedicated to the Pi [20]; in a wider sense, you may already know Stack Exchange, GitHub, Quora and others, but it is worth mentioning them in a little detail.

Stack Exchange, they say, has 173 communities built by people passionate about a focused topic [21]. Many of these are computer-related communities, or are otherwise scientific/technical in nature. You can just join and ask a question, and it may well be answered within minutes. It's also worth carefully trawling through old posts: if you're good at constructing Google queries, you can search something like "python iterate through list" and you'll get a ready-made chunk of code that allows you to go one by one down a list and do stuff. Indeed, such is the value of this approach that BioStars, founded as an online bioinformatics resource, was modeled after Stack Exchange [22]. Similar in spirit, GitHub is where one can host code, manage projects using the state-of-the-art distributed version-control system known as Git [23], and build software alongside millions of other developers. This is a central place to perhaps collaborate on projects, or maybe just download a pre-cooked module to run a Pi peripheral—for example, buy a 4-digit LED display (as for a clock radio), Google search "Python TM1637", and you'll find yourself at the doorstep of all the device drivers that some kind soul has already written and tested. Finally, Quora is a popular site too, modelled on a rich question-and-answer functionality and offering its own Pi-based communities.

Though sometimes a bit nerdy, you can't enter the above communities and not feel the warmth of human sharing, endeavour and altruism.

**Tip 8: It's versatile**

With 15m devices out there and peripherals costing literally pennies, it's unsurprising they find themselves everywhere. For a little inspiration, try the listicle "50 of the most important Raspberry Pi Sensors and Components" [24].

The Pi is also used across a wide range of sciences, from the amateur to the professional. For example, a webcam has been controlled by a Pi to monitor fiddler crabs [25], and a device known as the AirPi can record and upload information about temperature, humidity, air pressure, light levels, UV levels, carbon monoxide, nitrogen dioxide and smoke level to the internet [26]. If you want to go into the chemistry lab, then how about an Internet-based reaction monitoring, control and autonomous self-optimization platform for chemical synthesis [27]? And, if you're interested in the environment more generally, have a look at Public Lab—a community where you can learn how to investigate environmental concerns using inexpensive DIY techniques [28].

**Tip 9: You can do everyday science experiments with it for just pennies**

A friend wanted to know the least windy spot on his balcony as his plants were always getting blown over. A $5 anemometer rotor, a quick Google of the Hall Effect [29], some off-the-shelf-logging software, and the job was done. (It didn't stop his box hedge getting infested with caterpillars, but you can't have everything.)

Seriously though, most of the peripherals that you can get from online auction sites cost only a few dollars (yes, really); and, although most of them are somewhat rudimentary (e.g. measure distance ultrasonically to a few cm, light levels to a few lux), often they are good enough for basic, do-it-yourself (DIY) experimentation.

The Pi's ability to simply log data makes it tremendously useful, with only minimal expertise required. The Pi's GPIO pins essentially read if a voltage is there or not (they can read the level of a voltage as well, to an extent), so as long as you can make whatever you are running emit an electrical signal of some kind, then the Pi can log it (and it can also easily log radio frequency (Rf) signals with a tiny $5 superheterodyne receiver, read emails, check for HTTP or UDP messages, Bluetooth, and the list goes on). All of this makes interfacing with a Pi quite simple. In the direction of actuating (rather than detecting), the Pi is also capable of pulse-width modulation (PWM), which is a powerful technique for modulating digital signals. Closely linked to the idea of a 'duty cycle', PWM allows you to vary, in an analogue fashion, how much time the signal is in a high-intensity state; this, in turn, lets you dim LEDs, control the direction of a servo motor, and so on.

**Tip 10: You can do computational biology (and other science) with it, both teaching and research**

Phenomenally, given its affordability and simplicity, the Pi can be used for 'real' scientific pursuits, on both the research and teaching fronts. For example, a Pi has found its way into synthetic organic chemistry, where it has been the key hardware component in an integrated, open-source platform for the automated, machine-assisted synthesis of target compounds [30]. In that context, a Pi was used for real-time (online) data acquisition, analysis and feedback; by running control programs to orchestrate the activities of spectrometers, reactor units and so on, the Pi was the key module in a feedback loop for simultaneous control of multiple flow chemistry devices. Somewhat similar in spirit, a Pi Zero was recently used to control piezoelectric pumps in a "biocompatible, low-cost programmable dynamic flow pumping system"; dubbed 'PiFlow', this Pi-based system enables better mimicry of microphysiological environments [31]. Moving towards bioinformatics, 'PhyloPi' is a recent "affordable phylogenetic pipeline" that runs on a Pi and which is intended for use in HIV drug resistance testing facilities; the hardware and software engineering considerations that factored into this pipeline's design, including the Pi's limited memory (1GB RAM) and quad-core capabilities (enabling parallelized code to be run), are well-described in [32]. As shown in Figure 1, one can also generate publication-quality molecular illustrations, using a Pi and open-source code such as PyMOL [33], for less than $50 USD and about an hour of work, from start to finish.

On the teaching and pedagogy front, the Pi has been the centrepiece of significant efforts by Scottish research groups, based at Edinburgh and St Andrews, to develop and disseminate bioinformatics educational materials. These materials are freely distributed as an open-source project called '4273π' [34], which is also the name of the Raspbian-based OS image developed as part of the project.  Focused on bioinformatics and targeted at upper-level (fourth-year) biology undergraduates, the 4273π resource [35] provides one with training in basic Linux system and network administration, the Unix command-line interface, and so on—all invaluable skills in any computational science, biology and beyond.  In the same spirit, that research team has pioneered usage of the Pi as a way to teach the basics of computational sciences (quantitative and algorithmic thinking, programming principles [36]) to broader audiences, including high-school students [37] as well as the general public [38].  Finally, as an illustration of its general utility and versatility, note that the Pi is also being adopted for instructional purposes in other areas, such as in radiology training programs [39].

**Conclusion**

In summary, Raspberry Pi's are affordable, they are fun to use and they can be put to very serious uses too.  The skills and principles that you learn in getting them to work will serve you in good stead in today's world: you don't have to be (or become) a wizard at electronics, programming Python or even soldering, but rather just focus on gaining a practical understanding of what's going on; then, you can muddle through to make something useful. As in much of 'maker' and DIY culture [8], cosmetics is secondary to functionality; so, though most Pi projects might be in plastic sandwich boxes and held together with adhesive tape, they work, they're reliable, and their operation is transparent (no black boxes).

The world is changing rapidly, and you can only benefit from an understanding of the basics of technologies from the recent past, the present and the imminent future.  These technologies, current and future, include the Internet of Things, the principles behind autonomous and self-driving cars, many forms of pattern recognition (e.g. facial recognition), and, indeed, much of artificial intelligence—all of these will be part of the next generation's lives (and at least the tail-end of your own).  Most remarkably and promisingly, all of this new knowledge is accessible for the price of single family trip to the movies and some Googling. Indeed, one could argue that the educational role—and further potential—of the Raspberry Pi in democratizing and demystifying computing and technology is exceptionally valuable. The Pi's role in enabling primary scientific research is still at an early stage and is taking shape.

To end let us quote Francis Bacon, 1st Viscount St Alban, who around 1590 suggested "*scientia potestas est*" (*Knowledge is power*).  In the case of the Pi, this is roughly five volts at two amps, so it's ten watts. Frank– you were right on the money.

**Acknowledgments**

The authors thank Paul Rogers, John Royle, Deepak Winston and Professor Philip Bourne for review and comments. The work was partly supported by the Univ of Virginia and by U.S. National Science Foundation CAREER award MCB–1350957.

**Figure Captions**

**Figure 1**. A 'real' scientific example: molecular visualization with PyMOL on the Pi, in under an hour.  This montage shows a Pi being used to create publication-quality molecular graphics.  Starting from a pristine board (**A**), the Raspbian OS can be readily installed on, say, a 32GB MicroSD card such as shown in (**B**), via the New Out Of Box Software (NOOBS) installation manager; NOOBS is available at http://bit.ly/rasPiNOOB (which resolves to the longer URL in [40]).  Then, after installing PyMOL from the command-line and spawning a session (**C**), a protein 3D structure can be retrieved from the Protein Data Bank [41], loaded into PyMOL, and ray-traced (**D**); the PyMOL-related steps are detailed in (**E**), and primers such as [42] can be consulted for further details.  The approximate duration of each of the steps (**A**) → (**D**) is indicated alongside the panels.

**Figure 2**. Developing your own Pi-based system. This annotated photograph (taken by AC Fletcher) shows a typical Pi in action—under development, with an attached LCD display, audio cable, camera, network cable, 433-MHz radio transmitter and a buzzer.

Fletcher & Mura, **Figure 1**

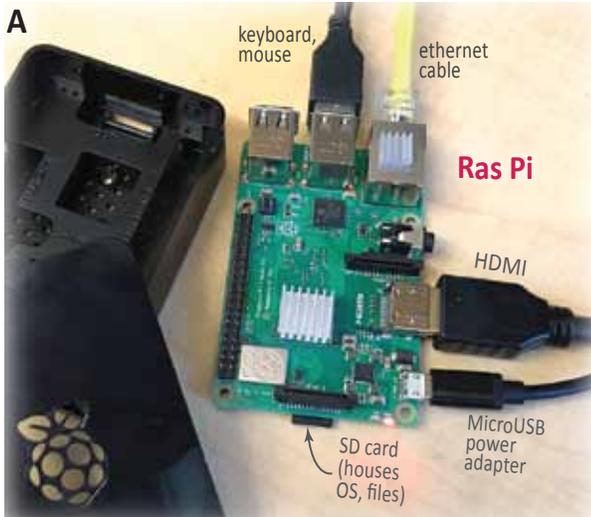

**A** Raspberry Pi 3 (Model B+); sample optional case (CanaKit™) shown at left; key interfaces labelled

**B** Installation of Raspbian OS (Debian-based Linux), from the MicroSD card (via NOOBS)

≈10 minutes → ≈15 minutes → ≈10 minutes → ≈5-10 minutes

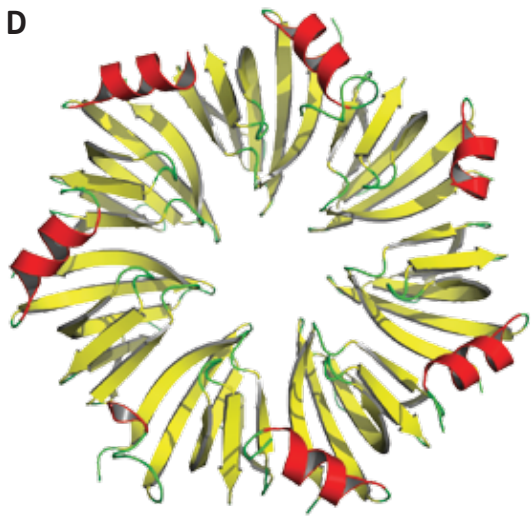

**D** Ray-traced rendering of a protein cartoon representation (PDB 1I8F)

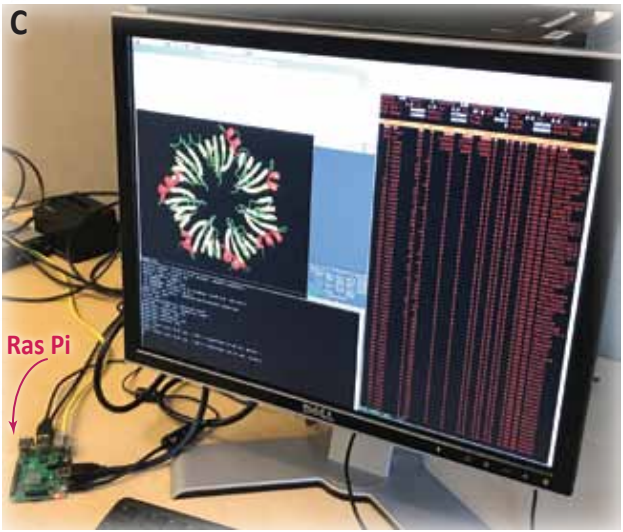

**C** PyMOL in action on the fresh Pi install; obtained via shell command "`sudo apt-get install pymol`"

**E** In the shell, type the command "`pymol`" (below), or navigate via desktop GUI menus (at right):

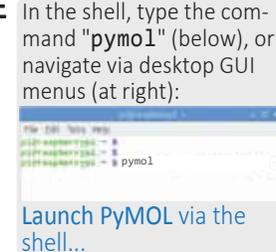

Launch PyMOL via the shell...   ...or via the RasPi desktop

Execute following commands at the PyMOL> prompt (press Enter key, ↵, after each):

```
PyMOL> fetch 1i8f
PyMOL> bg_color white
PyMOL> hide everything, all
PyMOL> show cartoon, all
PyMOL> util.cbss("1i8f","red","yellow","green")
PyMOL> ray
PyMOL> png myNicePic.png
```
Within PyMOL

Fletcher & Mura, **Figure 2**

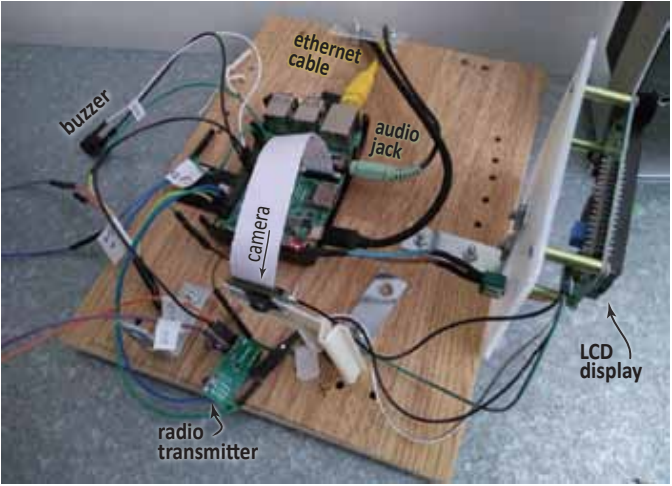